\documentclass[aps,pre,floatfix,nofootinbib,showpacs]{revtex4}
\usepackage{graphicx} 
\usepackage[sort&compress]{natbib}
\usepackage{graphicx}

\newcommand{\BEQ}{\begin{equation}}
\newcommand{\EEQ}{\end{equation}}
\newcommand{\BEA}{\begin{eqnarray}}
\newcommand{\EEA}{\end{eqnarray}}

\renewcommand{\k}{\alpha}
\renewcommand{\d}{{\rm d}}

\renewcommand{\l}{K}

\newcommand{\half}{\frac{1}{2}}

\newcommand{\CV}{{\cal V}}

\newcommand{\tk}{n}
\newcommand{\wphi}{\widetilde{\phi}}
\newcommand{\wpsi}{\widetilde{\psi}}

\begin{document}
\draft
\title
{How adsorption influences DNA denaturation}
\date{\today}

\author{A.E. Allahverdyan$^{1)}$, Zh.S. Gevorkian$^{1,2,3)}$,
Chin-Kun Hu$^{3,4)}$ and Th.M. Nieuwenhuizen$^{5)}$ }

\address{$^{1)}$Yerevan Physics Institute,
Alikhanian Brothers St. 2, Yerevan 375036, Armenia,\\
$^{2)}$Institute of Radiophysics and
Electronics, Ashtarak-2, 378410, Armenia,\\
$^{3)}$Institute of Physics, Academia of Sinica, Nankang, Taipei
11529, Taiwan\\
$^4$Center for Nonlinear and Complex Systems and Department of
Physics,
Chung-Yuan Christian University, Chungli 300, Taiwan,\\
$^{5)}$Institute for Theoretical Physics, University of Amsterdam,
Valckenierstraat 65, 1018 XE Amsterdam, The Netherlands
}

\begin{abstract}

The thermally induced denaturation of DNA in the presence of attractive
solid surface is studied.  The two strands of DNA are modeled via two
coupled flexible chains without volume interactions.  
If the two strands are adsorbed on the surface, the denaturation
phase-transition disappears. Instead, there is a smooth crossover to a
weakly naturated state.  Our second conclusion is that even when the
inter-strand attraction alone is too weak for creating a naturated
state at the given temperature, 
and also the surface-strand attraction alone is too weak for
creating an adsorbed state, the combined effect of the two attractions
can lead to a naturated and adsorbed state.  \end{abstract}


\maketitle

\section{Introduction}

The structure of DNA is the key for understanding its biological
functioning, explaining why the physical features of DNA have attracted
attention over the last decades
\cite{structure_function,freifelder,GKH,old_reviews,PB,Sung,PBreview}. A
known component of this structure is the Watson-Crick
double-strandedness: DNA is composed of two single-strand molecules
lined up by relatively weak hydrogen bonds. The double-strand exists for
physiological temperatures and is responsible for the stability of the
genetic information stored in DNA. For higher temperatures the
double-strand separates into two strands (denaturation).  Many processes
relevant for the functioning of DNA|such as transcription and
replication of the genetic information and packing of DNA into
chromosomes|proceed via at least partial separation (denaturation) of
the two strands due to breaking of hydrogen bonds
\cite{structure_function,freifelder}.  In addition, denaturation is
important for a number of technological processes, such as DNA sequence
determination and DNA mechanical nano-devices \cite{freifelder}. 

DNA denaturation is driven by changing the temperature or the solvent
structure, e.g., the pH factor \cite{structure_function,freifelder}.  There
are several generations of statistical physics models aiming to describe
the physics of denaturation. Early models, based on the one-dimensional
Ising model, focus on the statistics of hydrogen bonds modeling them
as two-state variables (open or closed) \cite{old_reviews}. More recent
models describe a richer physics, in that they try to explore space
configurations of DNA \cite{PB,Sung,PBreview,PB,Grass,kafri}.

Most of the physics literature devoted to DNA denaturation studies this
process in isolation from other relevant processes involved in DNA
functioning \cite{GKH,kwak,old_reviews,PB,Sung,PBreview,Grass,kafri}.
However, denaturation is frequently only a component of a larger
process, such as replication or compactification into a nucleosome, the
basic structural unit underlying the chromosome. 

Here we want to study how another important aspect of DNA physics
| adsorption of the double-strand DNA on a surface| influences its
denaturation.  Surface adsorption of DNA is widely employed in
biotechnologies for immobilization and patterning (drug or gene
delivery) of DNA \cite{dna_ad,kwak}. There are in fact several pertinent
situations, where both adsorption and denaturation of DNA are
simultaneously at play.

1) For DNA at normal conditions (pH$=7$ and NaCl concentration of $0.15$
M) thermal denaturation occurs between temperatures 67 C and 110 C
(which are the temperatures for A-T and C-G unbinding, respectively)
\cite{structure_function,GKH,old_reviews}. The denaturation temperature
can be decreased by increasing the pH factor, i.e., by decreasing the
concentration of free protons in the solvent, since the negatively
charged phosphate groups on each strand are not screened anymore by
protons and strongly repel each other. For the same reason, for the
DNA adsorption on a positively charged surface, the increase of the pH
will increase the electrostatic attraction to the surface. Thus at
certain values of the pH factor and the surface charge, denaturation and
adsorption may take place simultaneously.

2) Surface adsorption can be realized by the hydrogen-bonding of the
negatively charged phosphate residues to a negative surface (e.g.,
silica surface) \cite{melzak,kwak}. The effect is possible only when the
electrostatic repulsion is sufficienctly screened by the solvent
cations. Thus the same factors (temperature, pH, solvent concentration)
that decrease the inter-strand attraction, will weaken the DNA-surface
binding \cite{kwak}. 

3) The binding to hydrophobic surfaces (e.g., aldehyde-derivate glass,
or micro-porous membrane)
goes via partial denaturation which exposes the hydrophobic core of the
double-helix and leads to the DNA-surface attraction \cite{kwak}. Both
naturation and adsorption are simultaneously weakened by increasing the
pH \cite{kwak,tit}.

4) Human DNA has a total length of 2 m bearing a total charge of
$10^{8}$ electron charge units. This long object is contained in the
cell nucleus with diameter 10 $\mu$m, which is comparable with the
persistence length of DNA.  Recall that the persistence length of a
polymer is a characteristic length over which the polymer folds freely
due to thermal fluctuations. For the double-strand DNA at normal
conditions, the persistence length is relatively large and amounts to 50
nm or 100 base-pairs, while the persistence length of the
single-stranded DNA is much smaller, about 1-2 nm (i.e., 2-4 base-pairs)
\cite{ss_persistence}. This seems to create a paradoxical situation: not
only the large, strongly charged DNA has to be packed in a very small
compartment, but the DNA has to be replicated, repaired, and
transcribed. The problem is solved by a hierarchical structure: the DNA
double-helix is wrapped around positively charged histone (achieving
partial charge neutralization), histones condense into nucleosome
complex, which in its turn is contained in chromatin, {\it etc}. It was
recently discovered that packing of DNA into nucleosomes with
characteristic size much smaller than the persistence length of the DNA
chain proceeds via transient denaturation of the double strand
\cite{clotier}. Denaturation reduces the persistence length and thereby
facilitates the packing process. 

For all these processes we need to describe the DNA as a double-strand
polymer interacting with an attracting surface. This will be the goal of
the present paper. 

Needless to say that there is an obvious situation, where the
double-stranded structure is not relevant for the adsorption.  If the
two strands are too tightly connected, their separate motion is not
resolved. This case is well known in literature and|due to a large
persistence length of a double-strand DNA|can be described via an
effectively single semi-flexible chain interacting with the surface
\cite{singlestiff}.  These studies complement the classic theory of the
flexible chain adsorption, extensively treated in literature
\cite{GKH,deG,wiegel}. The electrostatic effects of the DNA adsorption,
modeled via a single Gaussian chain, are studied in \cite{joanny}.
Another recent activity couples the Ising-Zimm-Bragg model for the
helix-coil transition with the known theory of flexible chain adsorption
on solid surfaces \cite{carri}. While interesting for their own sake,
the results of Ref.~\cite{carri} do not apply to DNA
adsorption-denaturation, since the main assumption of
Ref.~\cite{carri}|that the helical pieces of the polymer interact with
the surface much stronger than the coiled ones|does not hold for DNA. 

This paper is organized as follows. In section \ref{section_model} we
define the model we shall work with. It describes two flexible chains
interacting with each other and with an attracting solid surface.
Section \ref{section_model} also recalls the known correspondence
between the equilibrium physics of flexible polymers and quantum
mechanics. In its final part this section discusses limitations of the
studied model in including volume interactions. Section
\ref{section_variational_principle} recalls the quantum mechanical
variational principle which will be the basic tool of our analysis.
Section \ref{section_no_melting} shows that if both polymers are
adsorbed on the surface, they do not denaturate via a phase-transition.
Section \ref{section_borromean_binding} discusses collective scenarios
of binding, while section \ref{section_no_binding} studies conditions
under which the naturated and/or adsorbed state is certainly absent. The
next section presents the phase diagram of the model. The last section
summarizes our results. Some technical issues are discussed in
appendices. The reader interested in the qualitative message of this
work may study section II for learning relevant notations and then jump
to section VII, which discusses general features of the phase diagram.
A short account of the present work has appeared already in
Ref.~\cite{letter}. 

\section{The model.}
\label{section_model}

When the motion of the single strands is resolved |i.e., when the
inter-strand hydrogen bonds are relatively weak, as happens next to
denaturation or unzipping transitions|DNA becomes a complex system with
different, mutually balancing features at play.  A realistic model of
DNA should take into account the stacking energy between two base pairs
and its dependence on the state (open or closed) of these pairs; helical
structure of the double-strand; intra-strand and inter-strand volume
interactions (e.g., self-avoidance); the pairing energy difference
between A-T and C-G pairs (respectively, $3\,k_{\rm B}T$ and $5\,k_{\rm B}T$
under normal conditions), etc.  Such fully realistic models do not seem
to exist; there are, however, various models with different degrees of
sophistication which are intended to capture at least some features of
the double-stranded structure
\cite{GKH,deG,singlestiff,PB,Sung,PBreview,PB,Grass,kafri}.

The model we shall work with disregards almost all the above complex
aspects and focuses on the most basic features of the problem.  It
consists of two homogeneous flexible chains interacting with each other
and coupled to the surface described as an infinite rigid attractive
wall.

Consider $2N$ coupled classical particles (monomers) with
radius-vectors $\vec{r}_{1| k}$ and $\vec{r}_{2| k}$
($k=1,...,N$) and potential energy
\begin{eqnarray}
\label{Ham}
\Pi(\vec{r}_{\alpha| k})=
\sum_{k=1}^N\left\{{\cal U}(\vec r_{k})+
\sum_{\alpha=1}^2
\left[\frac{\l}{2}
\left(\vec{r}_{\alpha| k}-\vec{r}_{\alpha| k-1}\right)^2
+{\cal V}(\vec{r}_{\alpha| k})
\right]\right\},
\end{eqnarray}
where $\vec r_{k}\equiv \vec r_{1| k}-\vec r_{2| k}$, so that
$|\vec r_{1| k}-\vec r_{2| k}|$
is the distance between two monomers, ${\cal U}$ is the
inter-strand potential and ${\cal V}$ is the surface-monomer potential.
The harmonic interaction with stiffness $\l$ (Gaussian chain)
between successive monomers in each strand is responsible for the
linear structure of the polymers. 

The system is embedded in an
equilibrium thermal bath at temperature $T=1/\beta$ ($k_{\rm B}=1$).
The quadratic kinetic energy of monomers is irrelevant, since it factorizes from the
partition function and does not influence the equilibrium probability distribution:
\BEA
\label{glum1}
&&P(\vec{r}_{\alpha| k})=
\frac{e^{-\beta \Pi(\vec{r}_{\alpha| k})}
}{{\cal Z}},\\
&&{\cal Z}=\int\left[\,\prod_{\alpha=1,2}\prod_{k=1}^N\, \d \vec{r}_{\alpha|k}\,\right]\,
e^{-\beta \Pi(\vec{r}_{\alpha| k})}.
\label{glum2}
\EEA

This model without adsorbing surface, i.e. ${\cal V}\equiv 0$, was
mentioned in \cite{deG} and studied in \cite{PB} in the context of DNA
denaturation. When the inter-strand interaction ${\cal U}(r)$ is absent,
we get two independent flexible chains interacting with the solid
surface, a well known model for adsorption-desorption phenomenon
\cite{GKH}. Recall that our purpose is in studying these two
processes|i.e., surface-polymer interaction and inter-strand
attraction|together. When taken separately, these processes are well
studied and well understood. 

Note that for the considered Gaussian chain model 
the stiffness parameter $\l$ relates to the characteristic persistence length $l_p$ as
\cite{GKH}:
\BEA
\label{LK}
\l=\frac{T}{l_p^2}.
\EEA

\subsection{Specification of the surface-monomer potential.}
\label{urchin1}

We assume that the surface can be represented as an infinite,
solid plane at $z=0$ (the role of the solid surface can be played by any body
of a smooth shape and the size much larger than the polymer length). 
Thus for the probability distribution
(\ref{glum1}) one has [for $\alpha=1,2$ and $k=1,\ldots,N$]
\BEA
P(\vec{r}_{\alpha| k})=
P(x_{\alpha| k}, y_{\alpha| k}, z_{\alpha| k})=0, \quad {\rm for}\quad
z_{\alpha| k}\leq 0.
\EEA
This boundary condition should be imposed as a constraint in (\ref{glum1}).

The remaining part of the surface-strand interaction is described by a
negative (attractive) potential ${\cal V}$ that depends only on the
third coordinate: ${\cal V}(\vec{r}_{\alpha| k})={\cal V}(z_{\alpha|
k})$. The potential ${\cal V}(z)$ will be assumed to be short-ranged: it
is negative for $z\to 0$ and tends to zero sufficiently quickly for
$z\to \infty$. 

Let us continue the specification of the potential ${\cal V}(z)$ taking
as an example the electrostatic attraction between one negatively
charged DNA strand and a positively charged surface; see, e.g.,
\cite{wiegel}.  We denote by $\sigma$ the surface charge density, $q$
stands for the monomer charge (for DNA the effective monomer charge is
roughly $q\simeq 1\,e$, where $e$ is the electron charge) and $\epsilon$
is the dielectric constant of the medium into which the polymer is
embedded ($\epsilon\approx 80$ for water at room temperature). Now the interaction energy
between the surface area $\d x\,\d y$ and one monomer reads:
\BEA
\label{laport}
\frac{q\sigma}{\epsilon r}\, e^{-k_{\rm D} r}\,\d x\,\d y,
\EEA
where $r=\sqrt{x^2+y^2+z^2}$ is the distance between the surface area $\d x\,\d y$ and the monomer,
while $k_{\rm D}$ is the inverse Debye screening length. This interaction leads to attraction for
opposite charges: $q\sigma<0$. 
The full expression of the inverse Debye screening length is well-known:
\BEA
\label{deb}
k_{\rm D}=\sqrt{2\pi l_{\rm B} \sum_a n_a Z^2_a}, \qquad
l_{\rm B}\equiv \frac{e^2}{\epsilon \,T},
\EEA
where $l_{\rm B}$ is the Bjerrum length, and where $n_a$ and $Z_a$ are,
respectively, the concentration and valency of ions of the sort $a$
present in the solvent (so that the ion charge is $Z_a e$).
The summation in (\ref{deb}) is taken over all sorts of ions present
\footnote{The quantity $\frac{1}{2}\sum_a n_a Z^2_a$ is called ionic strength. }.  Under
normal conditions the Bjerrum length is $\simeq 1$nm. At this length the
electrostatic interaction energy becomes comparable with the thermal
energy $T$.  The Debye length $1/k_{\rm D}$ varies between $\sim 0.5$nm and $\sim 1.5$nm 
under physiological conditions.  For pure water it is much longer: $1/k_{\rm D} \sim 1\mu$m.

Integrating (\ref{laport}) over $x$ and $y$ from $-\infty$ to $+\infty$, we get for the
surface-monomer interaction \cite{wiegel}:
\BEA
{\cal V}(z)= \frac{2\pi q\sigma}{\epsilon k_{\rm D}}\, e^{-k_{\rm D} z}.
\label{poto}
\EEA
Thus the strength of the potential is $\frac{2\pi q\sigma}{\epsilon
k_{\rm D}}$, while the inverse characteristic length is (expectedly) 
$1/k_{\rm D}$.  The potential ${\cal V}(z)$ is short-ranged for all other
relevant mechanisms of adsorption (hydrogen-binding, hydrophobic
interactions, cation exchange). This means, in particular, that
$\int_0^\infty \d z\,{\cal V}(z)$ is finite for all these mechanisms 
\cite{LL}. 

Returning to (\ref{poto}) we note that for a single flexible polymer
interacting with the surface the adsorption problem was solved in
Ref.~\cite{wiegel} within the Schr\"odinger equation approach to be
discussed below in detail; see in particular Eq.~(\ref{konrad_2}). 
The adsorption-desorption phase transition transition
temperature found in \cite{wiegel} reads:
\BEA
\label{tutu}
T_c=\frac{8.33\, \pi |\sigma q|}{k_{\rm D}^3 l_p^2 \epsilon},
\EEA
where $l_p$ is the persistence length from (\ref{LK}).

Let us estimate the Debye length as $k_{\rm D}^{-1}\sim 1$nm, the
single-polymer persistence length as $l_p\sim 1$nm, and assume that the
surface contains $Z$ elementary (electron) charges per $1\, {\rm nm}
\times 1\, {\rm nm}$.  Normally $Z\sim 1$, though stronlgy charged
surfaces achieve $Z=10-20$.  Taking the effective monomer charge one
elementary charge (which is a typical value for a single-strand DNA) and
recovering the Boltzmann constant, we see that (\ref{tutu}) predicts
$T_c$ of order of room temperature ($300$\,K). 

When looking at concrete parameters in (\ref{tutu}) we should also
recall that Eqs.~(\ref{laport}--\ref{poto}) account for the
surface-monomer electrostatic interaction, while the monomer-monomer
electrostatic interaction within the single polymer is neglected. This
is possible when the surface charge $\sigma l_p^2$ at the area
$l_p\times l_p$ (where $l_p$ is the persistence length of the single
strand) is larger than the monomer charge:
\BEA
\label{bala}
\sigma l_p^2 \gg |q|.
\EEA
This condition will be satisfied for strongly charged surfaces $Z\approx 10$.

\subsection{Specification of the monomer-monomer interaction between the two strands.}
\label{urchin2}

The inter-strand potential ${\cal U}(|\vec{r}_{1|k}-\vec{r}_{2|k}|)$
collects the effects of hydrogen-bonding, (partially) stacking, and
possible electrostatic repulsion. We again assume that it is purely
attractive, short-ranged and goes to zero sufficiently fast whenever the
inter-particle distance $|\vec{r}_{1|k}-\vec{r}_{2|k}|$ goes to
infinity. In particular, the short-ranged features implies that
$\int_0^\infty\d rr^2U(r)$ is finite. 

Several concrete examples of the inter-strand potential ${\cal U}$ were
studied and favorably compared with denaturation experiments in \cite{PB,Sung,PBreview}. 
For example, Ref.~\cite{Sung} studies the Morse potential
\BEA
\label{morse}
{\cal U}(r)=\nu e^{-a r} (e^{-a r}-2),
\EEA
where $\nu$ is the potential strength and $\frac{1}{a}$ is its characteristic range.
Within the Schr\"odinger equation approach [see (\ref{konrad_2}) below] 
Eq.~(\ref{morse}) predicts a second-order
denaturation transition at the critical temperature:
\BEA
\label{kuban}
T_c=\frac{16\nu}{a^2l_p^2}.
\EEA
Note that the appearance of the factor $a^2l_p^2$ in (\ref{kuban}) is
similar to the that of the factor $k_{\rm D}^2 l_p^2 $ in (\ref{tutu}).
Here are the standard estimates for the parameters in (\ref{kuban}):
$\nu\simeq 0.01$eV, $l_p\simeq 1$nm and $a\,l_p\simeq$2 \cite{PBreview}. These
produce from (\ref{kuban}) $T_c\sim 400$K, which by the order of
magnitude coincides with experimental values \cite{PBreview}.

\subsection{Effective Schr\"odinger equation.}
\label{urchin3}

It is known, see e.g. \cite{GKH,PBreview}, that in the thermodynamical
limit $N\gg 1$ the free energy of flexible polymer in an external
potential is determined from an effective Schr\"odinger equation; see
Appendix \ref{effective_wave_equation} for more details. A sufficient
condition for validity of the Schr\"odinger equation approach is that
the characteristic length $D$ over which the polymer density changes
is much larger than the persistence length $l_p$: 
\BEA
\label{bahatur}
D\gg l_p. 
\EEA
This
condition is always satisfied in the vicinity of a second-order
phase-transition, where $D$ is naturally large for a fixed $l_p$. 
If condition (\ref{bahatur}) is satisfied for a short-range potential|see
(\ref{poto}, \ref{kuban}) for relevant examples|this potential is necessarily 
small for those distances $\sim D$, where
the flexible polymer is predominantly located \cite{GKH}.

For the considered two-strand situation
the Schr\"odinger equation reads
\BEA
\label{1a}&& H\Psi=E\Psi,\\
\label{2a}&&H\equiv{\sum}_{\k=1}^2[
-\half\,\partial_{\vec{r}_{\k}}^{\,2}
+ V(z_\k)]+U(r),
\EEA
where [using also (\ref{LK})]
\BEA
V(z)\equiv \l\beta^2\,{\cal V}(z)=\frac{\beta}{l_p^2}
\,{\cal V}(z),\qquad U(r)\equiv \l\beta^2\,{\cal U}(r)=\frac{\beta}{l_p^2}\,{\cal U}(r).
\EEA

If there is a gap between the lowest two eigenvalues of $H$, the
ground state wave-function $\Psi$ determines the monomer
statistics as \BEA n(\vec{r_1},
\vec{r_2})=\Psi^2(\vec{r}_1,\vec{r}_2), \EEA where $n(\vec{r_1},
\vec{r_2})$ is the probability distribution for two neighboring
monomers on the strands for the considered translationally
invariant system.

Recalling the known correspondence between the flexible polymer physics
and (stationary) quantum mechanics \cite{GKH}, we can think of
$\vec{r}_{1,2}=(x_{1,2},y_{1,2},z_{1,2})$ as the position vectors of two
quantum particles representing the strands, while
$\vec{r}=\vec{r}_1-\vec{r}_2$ is their mutual position. 

The eigenvalue $E$ is the energy of the quantum pair. It is related
to the free energy $fN$ of the system as
\BEA
E=\beta^2 l f+3\beta l\ln \frac{2\pi}{\beta l}.
\EEA

Since the surface is described by an infinite potential wall,
we have the following boundary condition for the wave function
\footnote{In fact, one should be more careful, when defining the
  boundary condition (\ref{4}). For the two-particle case it appears
  to be necessary to fix not only the continuity of $\Psi$ and its
  value at the wall, as Eq.~(\ref{4}) does, but also the behavior next
  to the wall: one has to require that when $z_1$ and $z_2$ go to
  zero simultaneously, $\Psi\propto z_1z_2$. Otherwise, there will be (continuous)
  wave-functions which provide a bound state for two-particles with an
  arbitrarily weak $V<0$ and arbitrary weak inter-particle
  interaction $U<0$, though the single particle needs a critical
  strength of $V$ to get into a bound state. This obviously
  pathological situation is prevented by the additional boundary
  condition $\Psi\propto z_1z_2$.  For the wave-functions we shall consider below this
  additional boundary condition will be satisfied automatically.  }
\BEA
\label{4}
\Psi(\vec{r}_1,\vec{r}_2)=0, \quad {\rm if}\quad z_1\leq 0, \quad
{\rm or}\quad z_2\leq 0.
\EEA

Both $V(z)$ and $U(r)$ are attractive, $V\le 0,\,U\le 0$, and
short-ranged, that is $\int_0^\infty\d zV(z)$ and $\int_0^\infty\d
rr^2U(r)$ are finite.  When $U=0$, the Hamiltonian $H$ reduces to two
uncoupled strands (or two uncoupled quantum particles), each one in the
potential $V(z)$. The corresponding Schr\"odinger equation for the $z$-coordinate 
of one strand reads from (\ref{2a}):
\BEA
\label{konrad_1}
[-\frac{1}{2}\partial_z^2+V(z)]\psi(z)=E\psi(z), \qquad \psi(z=0)=0.
\EEA

It is well-known that if $V(z)$ is shallow enough, no bound
(negative energy) state exists, while the second-order binding
transitions corresponds to adsorption of a single flexible polymer
\cite{GKH}. The physical order-parameter for this transition is the
inverse square average distance from the surface, $1/\langle
z^2\rangle$, which is finite (zero) in the adsorbed (desorbed) state.
It is useful to denote by $\mu$ the dimensionless coupling constant of
$V=\mu\widetilde{V}$ such that (for $U=0$) the adsorption threshold is
\BEA
\label{petan}
\mu_{c,0}=1.
\EEA
Note that the adsorption of a single strand DNA is a part of the renaturation
via hybridization \cite{freifelder}, a known method of genetic systematics.

For the example (\ref{poto}) the concrete expression for $\mu$ reads from (\ref{tutu}):
\BEA
\label{eth1}
\mu=\frac{8.33\, \pi |\sigma q|}{T\,k_{\rm D}^3 l_p^2 \epsilon}.
\EEA

Analogously, switching off both $V(z)$ and the wall, we shall get
a three-dimensional central-symmetric motion in the potential
$U(r)$ which again is not bound if $U$ is shallow. This
second-order unbinding transition with the order parameter
$1/\langle r^2\rangle$, where $r$ is the inter-strand distance,
corresponds to thermal denaturation (strand separation) of the
double-strand polymer
\cite{structure_function,freifelder,PB,PBreview}. 

The Schr\"odinger equation for the radial motion in the absence of the surface
reads from (\ref{2a}) \cite{PBreview}
\BEA
\label{konrad_2}
[-\frac{1}{4}\partial_r^2+U(r)]\chi(r)=E\chi(r), \qquad \chi(r=0)=0,
\EEA
where $\chi(r)$ is related to the original wave-function as
\BEA
\psi(r)=\frac{\chi(r)}{r}.
\EEA

Note that (\ref{konrad_2}) is again a one-dimensional Schr\"odinger equation, but
as compared to the equation (\ref{konrad_1}), Eq.~(\ref{konrad_2}) contains an additional factor
$\frac{1}{2}$ next to the kinetic-energy term $\partial_r^2$. This factor arises due to effective
mass; see \cite{LL} for more details.

Let us write likewise
$U=\lambda \tilde U$, where $\lambda$ is the dimensionless naturation
strength. We take the naturation threshold in the bulk to be 
\BEA
\label{mukhan} \lambda_{c,0}=1. 
\EEA 
For the example (\ref{morse}), $\lambda$ reads from (\ref{kuban}):
\BEA
\label{eth2}
\lambda=\frac{16\nu}{a^2l_p^2T}.
\EEA

When the wall is included, i.e., condition (\ref{4}) is imposed, the
strands loose in the adsorbed phase part of their entropy. This is known
to lead to a fluctuation induced effective repulsion \cite{LN}. 

Let us now remind that the physics of weakly bound quantum particles
does not depend on details of binding potential \cite{LL}. Thus for
qualitative understanding of the situation one may employ the
delta-shell potential, which is easily and exactly solvable and has very
transparent physical features; see Appendix \ref{apa}.

\subsection{Relevant coordinates.}

Let us now return to the basic equation (\ref{konrad_1}).
It is convenient to recast this equation in new coordinates:
\BEA
&&v_1=\half(x_1+x_2),\qquad
v_2=\half(y_1+y_2),\\
&&x_1-x_2=\rho\cos\varphi,\qquad
y_1-y_2=\rho\sin\varphi,
\EEA
where
\BEA
0\leq\rho,\qquad 0\leq \phi\leq 2\pi,
\EEA
and to re-write the Schr\"odinger equation
(\ref{1a}, \ref{2a}) as
\BEA
\label{2b}
-\frac{1}{2}\left \{
\frac{2}{\rho}
\frac{\partial}{\partial \rho}
\rho\frac{\partial}{\partial \rho}+
\frac{1}{\rho^2}\frac{\partial}{\partial \varphi}
+\frac{1}{2}\frac{\partial^2}{\partial v_1^2}
+\frac{1}{2}\frac{\partial^2}{\partial v_2^2}
+\frac{\partial^2}{\partial z_1^2}
+\frac{\partial^2}{\partial z_2^2}
\right\}\Psi+\left\{
V(z_1)+V(z_2)+U(|\vec{r}_1-\vec{r}_2|)\right\}\Psi
=E\Psi.
\EEA

It is seen from (\ref{2b}) that the variables separate, since
$\Psi(\vec{r}_1,\vec{r}_2)$ can be written as
\BEA
\Psi(\vec{r}_1,\vec{r}_2)=\psi(\rho,z_1,z_2)\psi_1(v_1)\,
\,\psi_2(v_2)\,\psi_3(\varphi),
\EEA
and the lowest energy levels is to be found via
the following equation
\BEA
\label{3}
-\frac{1}{2}\left \{
\frac{2}{\rho}
\frac{\partial}{\partial \rho}
\rho\frac{\partial}{\partial \rho}
+\frac{\partial^2}{\partial z_1^2}
+\frac{\partial^2}{\partial z_2^2}
\right\}\Psi+\left\{
V(z_1)+V(z_2)+U(\sqrt{\rho^2+(z_1-z_2)^2})\right\}\Psi
=E\Psi.
\EEA

Thus due to the translational invariance along the surface and the
invariance under rotations around the $z$-axis, we are left with
three independent coordinates: the projection $\rho$ of the
inter-particle distance on the surface, and the distances $z_1$,
$z_2$ between the particles and the surface.

Note that within the quantum mechanical setting the problem described by (\ref{3})
corresponds  to a three-body problem, where the role of the third body (with infinite mass)
is played by the surface.

\subsection{Common action of the surface-strand and inter-strand potentials.} 

In this and subsequent subsection we shall discuss two possible limitations
of the present model.

Above we combined together the surface-strand interaction potential
${\cal V}$, which was derived separately from studying interaction of
the surface with one flexible strand, and inter-strand potential ${\cal
U}$ deduced from studying two flexible strands without the surface.
While this type of combining is widely applied in all areas of
statistical physics, its applicability needs careful discussions in each
concrete case.  For instance, it is possible that the presence of
adsorbing surface will directly influence the inter-strand potential.
Let us discuss one (perhaps the major) example of that type pertinent
for the studied model. 

It is well-known that the two strands of DNA are negatively charged
\cite{structure_function}.  For the double-stranded DNA under normal
conditions the inter-strand repulsion is screened by positive counterions, so that
the hydrogen bonding can overcome the electrostatic repulsion and create
an effective attraction, which is then the main reason of inter-strand
binding \cite{structure_function}. Once DNA denaturates and separates
into two strands, the counterions are released into the ambient medium
and are clouded around each strand. However, for temperatures not very
far from the denaturation temperature the counterions continue to screen
the electrostatic repulsion, so that once the temperature lowers below
the denaturation transition temperature, the two strands reversibly
assemble back into the double-strand \cite{structure_function}. We
stress that the fact that (partially released) counterions still provide
a sufficient screening follows from the existence of the observed reversible
renaturation transition. 

When DNA denaturates in the presence of a positively charged surface the
cloud of screening counterions around each strand will tend to rarefy.
This will increase the screening length and make the overall
inter-strand interaction repulsive. However, this is possible only for
strongly adsorbed strands, where the majority of counterions are within
the direct influence of the surface charge. In the present work we focus
on weakly bound strands, where the characteristic length of the adsorbed
layer $D$ is much larger than the persistence length $l_p$
(approximately $1$nm in normal conditions), which is of the same order of magnitude as
the Debye screening length $1/k_{\rm D}$;
see (\ref{bahatur}). Thus the majority of counterions will not feel the
adsorbing surface, and in this case we do not need to account directly
for the influence of the surface on the inter-strand potential.  For
strongly adsorbed DNA strands, i.e., for $D\sim 1/k_{\rm D}$, it can be
necessary to couple directly the inter-strand potential with the degree
of adsorption.

\subsection{Self-avoidance and of electrostatic volume interactions.}
\label{tartar}

In Hamiltonian (\ref{Ham}) we accounted for the surface-strand and
inter-strand interaction, but neglected all the volume interactions such as
self-avoidance and (for charged polymers) electrostatic interaction
between various monomers. It is important to note that the volume interactions
coming from the intra-chain contributions can be accounted for within the present model via
renormalizing the persistence length $l_p$; see (\ref{Ham}) and
(\ref{LK}) for definitions. As shown in \cite{mut} for a single flexible
polymer interacting with electrostatically adsorbing surface, the
self-avoiding interactions and electrostatic volume interactions
renormalize the persistence length. Provided that the Debye screening
length $1/k_{\rm D}$ is not very large |a sufficient condition for this
is $k_{\rm D}l_p\sqrt{N}\gg 1$, where $N$ is the number of monomers
\cite{mut}| both self-avoiding and electrostatic volume interactions
lead to an effective persistence length $\widetilde{l}_p$, which differs
from the bare persistence length mainly by the factor $N^{1/10}$:
$\widetilde{l}_p\sim N^{1/10}\, l_p$ \cite{mut} (the remaining part of
renormalization is numerical factors, which are not essential for the
present qualitative discussion) \footnote{In more detail,
Ref.~\cite{mut} considers a continuous polymer model with length $L$ and
reports for the square of the effective persistence length
$\widetilde{l}_p^2\sim L^{1/5}\, l_p^2$. For the present dsicrete model
we take naturally $L\propto N$.}. Once the persistence length is
renormalized, one can still use the flexible polymer coupled to an
adsorbing surface \cite{mut}. Thus the transition temperatures
(\ref{tutu}) and (\ref{kuban}) are divided by factor $N^{1/5}$, where
$N$ is the number of monomers. Now for the typical single-strand DNA
length $N\sim 10^4$ this renormalization will not make any
substantial change in transition temperatures, though it is essential
for longer polymers, $N\geq 10^5$. In particular, for such a long
polymer the persistence length may increase to an extent that the
condition (\ref{bahatur}) will be violated. 

We will see below that for qualitative conclusions of this paper, the
precise form of the renormalized persistence length is not essential,
provided that one can still employ the Schr\"odinger equation
(\ref{konrad_1}) for describing denaturation and desorption. The main
reason for this is that the renormalization of the persistence length
homogeneously renormalizes both dimensionless couplings $\lambda$ and
$\mu$ in (\ref{mukhan}) and (\ref{petan}), respectively. 

The above discussion does not account for the inter-chain volume
interactions and thus should not create an impression that the full
volume interactions effect for two coupled chains can be described via a
renormalized persistence length. It is clear that one needs a more
specific study of volume interactions for the present model. Since such
a study poses immence analytical problems, it will be concluded at a
later time.

\section{Variational principle and the existence of the
overall bound states.}
\label{exi}
\label{section_variational_principle}

Note that Eqs.~(\ref{3}, \ref{4}) follow from a variational
principle:
\BEA
\label{var0}
\delta {\cal I}\{\psi\}=0,
\EEA
with
\BEA
&&{\cal I}\{\psi\}=
\int_0^\infty\int_0^\infty\int_0^\infty
\rho\d \rho\,\d z_1\d z_2\nonumber\\
&&\left[
\frac{1}{4}\left\{
2\left( \frac{\partial\psi}{\partial \rho} \right)^2+
\left( \frac{\partial\psi}{\partial z_1} \right)^2+
\left( \frac{\partial\psi}{\partial z_2} \right)^2
\right\}+\frac{1}{2}\left\{
V(z_1)+V(z_2)+U(\sqrt{\rho^2+(z_1-z_2)^2})-E
\right\}\psi^2
\right]=0,
\label{var}
\EEA
where $\psi$ is taken real, since we are interested in bounded
(discrete-level) states. We already assumed that $\psi$ is properly
normalized:
\BEA
\int_0^\infty\int_0^\infty\int_0^\infty
\rho\d \rho\,\d z_1\d z_2\,\psi^2(z_1,z_2,\rho)=1.
\EEA

If either $V(z)=0$ or $U=0$, the criterion for the existence of a
bound state is well known, since it reduces to the existence of a
negative energy in the spectrum, or equivalently to the existence of a
physically admissible (satisfying the proper boundary conditions)
wave-function with a negative average energy.

The situation is slightly more delicate when the two potentials $V$ and
$U$ act together. Let us assume that
either for $V(z)\to 0$ or for $U(|\vec{r}_1-\vec{r}_2|)\to 0$
there are negative energy states. Denote by
\BEA
E\{U\}<0,\qquad
E\{V(z_1)+V(z_2)\}=2E\{V\}<0,
\EEA
respectively, the corresponding lowest (most negative) energies.

Then it suffices to have a normalized
wave-function $\psi$ with
\BEA
\label{ter2}
{\cal I}\{\psi\}<
E\{U\}+2E\{V\},
\label{amiak1}
\EEA
for at least one overall bound, i.e., adsorbed and naturated, state to exist.

\section{Absence of denaturation phase-transition for adsorbed strands.}
\label{section_no_melting}

Let us return to the variational principle (\ref{var}) and assume that $V(z)$
is strong enough to create at least a single (lowest) bound state with
energy $E\{V\}<0$. Denote by $\phi(z)$ the corresponding lowest-energy
normalized wave function:
\BEA
-\frac{1}{2}\phi''(z)+V(z)\phi(z)=E\{V\}\,\phi(z).
\EEA

For the overall problem we shall employ the following
variational wave-function:
\BEA
\label{ppt}
\psi(\rho,z_1,z_2)=\phi(z_1)\,\phi(z_2)\,\xi(\rho),
\EEA
where $\xi(\rho)$ is an unknown, tentatively normalized, viz.
\BEA
\label{tent}
\int_0^\infty \d
\rho\rho \xi^2(\rho)=1,
\EEA
wave-function, to be determined from the
optimization of (\ref{var}). Note that in (\ref{ppt}) the boundary
conditions for the surface are satisfied via $\phi(z_1)\,\phi(z_2)$.

Substituting (\ref{ppt}) into (\ref{var}) and varying it over $\xi$,
we get an effective Schr\"odinger equation for $\xi(\rho)$:
\BEA
\label{303}
-\left \{
\frac{1}{\rho}
\frac{\partial}{\partial \rho}
\rho\frac{\partial}{\partial \rho}
\right\}\xi+\left\{
U_{\rm eff}(\rho)
-\varepsilon\right\}\xi
=0,
\EEA
where $U_{\rm eff}(\rho)$ is an effective potential:
\BEA
\label{kokand}
U_{\rm eff}(\rho)=\int_0^\infty\d z_1\int_0^\infty\d z_2\,
\phi^2(z_1)\,\phi^2(z_2)\,
U(\,\sqrt{\rho^2+(z_1-z_2)^2}\,),
\EEA
and where $\varepsilon$ is the reduced energy
\BEA
\varepsilon=E-2E\{V\}.
\EEA

Two main point about the effective potential (\ref{kokand}) is
that it is attractive (since so is $U$) and goes to zero for
$\rho\to\infty$. The last feature follows from the analogous one
of $U(r)$ and the fact that $\phi(z)$ are normalizable. A more
explicit form for $U_{\rm eff}$ can be obtained by assuming that
$U(r)$ is a delta-shell potential \BEA \label{ken}
U(r)=-\frac{\lambda}{r_0}\,\delta(r-r_0), \EEA with the strength
$\lambda>0$ and the attraction radius $r_0>0$. The transparent
properties of this potential are recalled in Appendix \ref{apa}.
The critical binding strength of this potential is \BEA
\label{baran} \lambda_{c,0}=1, \EEA as given by (\ref{crit}). [When
comparing Eq.~(\ref{ken}) with Eq.~(\ref{kuma}), note that the
additional factor $2$ comes from the reduced mass.]

Using (\ref{ken}) we now obtain from (\ref{kokand}) after changing
variables:
\BEA
U_{\rm eff}(\rho)=&&-\frac{\lambda}{r_0}\,\,
\int_0^\infty\d v\int_0^v\d u\,\phi^2\left(\frac{v+u}{2}    \right)
\, \phi^2\left(\frac{v-u}{2}    \right)\,\delta\left(
\sqrt{\rho^2+u^2}-r_0
\right)\nonumber\\
=&&-\lambda\,\,
\frac{\theta(r_0-\rho)}{\sqrt{r_0^2-\rho^2}}
\int_{\sqrt{r_0^2-\rho^2}}^\infty\d v\,
\phi^2\left(\frac{v+\sqrt{r_0^2-\rho^2}}{2}\right)
\,\phi^2\left(\frac{v-\sqrt{r_0^2-\rho^2}}{2}\right)
\nonumber\\
=&&-2\lambda\,\,
\frac{\theta(r_0-\rho)}{\sqrt{r_0^2-\rho^2}}
\int_0^\infty\d v\,
\phi^2\left(v+\sqrt{r_0^2-\rho^2}\right)\,\phi^2(v).
\label{artashir}
\EEA
It is now seen explicitly that $U_{\rm eff}(\rho)$ is zero for
sufficiently large $\rho$.

Note that (\ref{303}) has the form of two-dimensional Schr\"odinger
equation for an effective particle in the attractive potential $U_{\rm
  eff}(\rho)$. It is well known that any (however weak) attractive
potential in two dimensions creates a bound state \cite{LL}.
Thus there is a
normalizable function $\xi(\rho)$ such that $\varepsilon$ in
(\ref{kokand}) is negative. This means that
\BEA
E<2E\{V\},
\EEA
and, according to our discussion in section
\ref{section_variational_principle}, there is an overall bound
(naturated and adsorbed) state provided $V(z)$ creates a bound state. In
our model a sufficiently attractive surface potential confines
fluctuations of the two strands and prevents the denaturation
phase-transition (this however does not mean that the denaturation is
absent as a physical process; see below).

The physical reason for the existence of an overall bound state
for an arbitrary small potential is a peculiar two-dimensional
effect: the weakly singular attractive $\propto 1/\rho^2$
potential \cite{LN} \footnote{One-dimension in this respect is not much
different from the three-dimensional situation. The known
statement on the existence of bound state for any small
one-dimensional potential is connected with a different mechanism,
that is, with allowing all values of the one-dimensional
coordinate (no infinite wall at the origin). The two-dimensional
situation is indeed peculiar in this respect.  }.  Indeed changing
in (\ref{303}) the variables as 
\BEA \label{304} \widetilde{\xi}
=\frac{\xi}{\sqrt{\rho}}, 
\EEA 
we get 
\BEA \label{305}
-\frac{\partial^2\widetilde{\xi}}{\partial^2 \rho} +\left\{ U_{\rm
eff}(\rho)-\frac{1}{4\rho^2} 
-\varepsilon\right\}\widetilde{\xi}
=0. 
\EEA 

Eq.~(\ref{304}) implies \BEA \widetilde{\xi}(0)=0, \EEA
i.e., the existence of the infinite wall at $\rho=0$ for the
effectively one-dimensional Eq.~(\ref{305}). It is, however, seen
from (\ref{305}) that there is also an attractive potential
$1/(4\rho^2)$. It is known that if the strength of such a potential
is larger than $1/4$ seen in (\ref{305}), the (effective) quantum
particle will fall to zero, i.e., the ground state will be minus
infinity \cite{LL}.  The value $1/4$ is just at the border of this
phenomenon and, therefore, any attractive short-range potential
acting in addition to $1/(4\rho^2)$ suffices to create a bound state \cite{LN}.

To illustrate the behavior of $U_{\rm eff}$
for weakly bound state of the potential $V(z)$,
let us assume that $V(z)$ is also
a delta-shell potential:
\BEA
\label{kendo}
V(z)=-\frac{\mu}{2z_0}\,\delta(z-z_0).
\EEA
Recall that we still have an infinite wall at $z=0$
and that for the delta-shell
potential the bound state exists for
\BEA
\label{dur}
\mu> \mu_{c,0}=1,
\EEA
see Appendix \ref{apa} for details. If now $\mu$ is close to one,
the energy $E\{V\}\equiv -k^2/2$ is small.
Working out (\ref{artashir}) with help
of Eq.~(\ref{asa}), which essentially reduces to
\BEA
\label{asa1}
\phi(z)\propto \sqrt{2k}\,e^{-kz},
\EEA
we get
\BEA
\label{uralmash}
U_{\rm eff}(\rho)
=-2\lambda k\,\frac{\theta(r_0-\rho)}{\sqrt{r_0^2-\rho^2}}
\EEA
Since for small $k$, the wave-function $\psi(x)$ is almost
delocalized, the effective potential $U_{\rm eff}(\rho) $ is
proportional to $k$ and goes to zero for $k\to 0$ that is for $\nu\to
1$. In other words, the trial function (\ref{ppt}) does not predict
any (overall) binding for
\BEA
\mu\leq 1.
\EEA

Note however that although for $\mu>1$ any inter-strand attraction is
able to prevent the denaturation phase transition, the energy
$\varepsilon$ in (\ref{305}) is exponentially small for small $U_{\rm
eff}$, i.e., small $\lambda$ or small $k$. Recall that this energy is estimated
as \cite{LL}
\BEA
\varepsilon \simeq \frac{2}{r_0^2}\exp\left[
2\int_0^\infty \d \rho \rho U_{\rm eff}(\rho)
\right]=\frac{2}{r_0^2}\, e^{-1/(\lambda k r_0)}.
\EEA

Thus for a small $\lambda$ or $k$ we get a very 
large separation between the strands. In this sense
the (incomplete) denaturation phenomenon without the phase transition
is present in our model. 

In summary, the main physical message of this section is that if the two
strands are localized near the surface, the overall DNA molecule does
not melt via a phase-transition with increasing the temperature: there
is only a smooth crossover from tightly bound to a (very) weakly bound
state. The cause of this effect is that the surface confines
fluctuations of each strand. Mathematically this is expressed by an
additional attractive potential $-\frac{1}{4\rho^2}$ in (\ref{305}). 

This result was obtained without taking into account various realistic
features of DNA. It is possible that the denaturation transition in the
adsorbed phase will recover upon taking into account some of those
neglected features, e.g., volume interactions between the two strands
and within each strand (see \cite{kafri} for a prediction of such a
transition in a different model of DNA that partially accounts for
volume interactions).

We nevertheless expect that the obtained result
will apply, at least qualitatively, to denaturation-renaturation
experiments, and will be displayed by facilitation of the naturation in
the adsorbed phase. We are not aware of any specific experiment done to
check the renaturation-facilitating effect of an attractive surface.
There are, however, somewhat related experiments showing that the
renaturation rate can significantly increase in the condensed (globular)
phase of single-strand DNA \cite{sikorav}. This condensed phase is
created by volume (monomer-monomer) interactions. The effect was
obtained under rather diverse set of conditions, but to our knowledge it
did not get any unifying explanation. The analogy with our finding is
that in the condensed phase fluctuations of the single strand DNA are
also greatly reduced as compared to coil (free) state.

\section{Collective binding.}
\label{section_borromean_binding}

With the aim to understand the situation when $V(z)$ alone does not provide any binding,
we take for the variational function
\BEA
\psi(z_1,z_2,\rho)=\phi(z_1,z_2)\,\xi(\rho).
\label{ppp}
\EEA
As compared to (\ref{ppt}) we do not require that $z_1$ and $z_2$ are
factorized, and we are going to optimize over $\phi(z_1,z_2)$. In
contrast, $\xi(\rho)$ is a fixed, normalized (see (\ref{tent})\,) known
function.

Substituting (\ref{ppp}) into (\ref{var}) and varying over
$\phi(z_1,z_2)$ we get:
\BEA
\label{oman}
-\frac{1}{2}\left \{
\frac{\partial^2}{\partial z_1^2}
+\frac{\partial^2}{\partial z_2^2}
\right\}\phi+\left\{
V(z_1)+V(z_2)+V_{\rm eff}(|z_1-z_2|)
-E_1\right\}\phi
=0,
\EEA
where
\BEA
\label{katar}
V_{\rm eff}(z)
\equiv\int_0^\infty\d \rho\,\rho\,
\xi^2(\rho)\,U(\,\sqrt{\rho^2+z^2}\,).
\EEA

Recall that by the very meaning of the variational approach $E_1$
provides |for any $\lambda$ and any normalized function $\xi(\rho)$|
an upper bound for the real ground state energy.
Eq.~(\ref{oman}) describes two one-dimensional particles
with inter-particle interaction $V_{\rm eff}(|z_1-z_2|)$
and coupled to an external field $V(z)$.

For the inter-particle interaction given as in (\ref{ken}), this
effective potential $V_{\rm eff}(z)$ reads
\BEA
\label{ulf}
V_{\rm eff}(z)
=-\lambda\,\theta(r_0-z)\,\xi^2\left(\sqrt{r_0^2-z^2}\right).
\EEA

We are now going to show that Eq.~(\ref{oman}) predicts binding
|that is, it predicts $E_1<0$ and a localized normalizable
wave-function $\phi(z_1,z_2)$| at the critical point $\mu=1$ of
the potential $V(z)$.  To this end let us calculate the
perturbative correction $\Delta E$ introduced by the effective
potential $V_{\rm eff}$. At first
  glance the application of perturbation theory is problematic, because
  we search for a nearly degenerate energy level. However, due to strong
  delocalization of the corresponding wave-function, the matrix
  elements of the perturbing potential $V_{\rm eff}$ appear to be
  small as well, and applying perturbation theory is legitimate. This
  will be also underlined below by a perfectly finite behavior of the second
  order perturbation theory result.

Recall that in the first two orders of the perturbation theory we have
\cite{LL}
\BEA
\label{gamow}
&&\Delta E\equiv
E_1-2E\{V(z) \}
=\langle 0|V_{\rm eff}|0\rangle -\int_0^\infty\d K\,
\frac{|\langle 0|V_{\rm eff}|nK\rangle  |^2}{\varepsilon_K-2E\{V \}},\\
&& 2E\{V \}=-k^2,
\EEA
where $\langle z_1,z_2|\,0\rangle =\phi(z_1)\phi(z_2)$ is the lowest energy state of
the unperturbed system, and where the integration over $K$ involves
all excited wave-functions of the unperturbed two-particle system with
wave-vector $K$ and energy $\varepsilon_K$ (all these wave-functions
are in the continuous spectrum). Note that there
are three orthogonal families of these states:
\BEA
\label{hop1}
&&\phi(z_1)\wphi(nz_2,n), \qquad\qquad \varepsilon_n=
\frac{n^2}{2}-\frac{k^2}{2},
\\
\label{hop2}
&&\phi(z_2)\wphi(nz_1,n), \qquad\qquad \varepsilon_n=\frac{n^2}{2}
-\frac{k^2}{2},
\\
\label{hop3} &&\wphi(nz_1,n_1)~\wphi(nz_2,n_2), \qquad
\varepsilon_{n_1\,n_2}= \frac{n_1^2}{2}+ \frac{n_2^2}{2}, \EEA
where $\wphi(nz,n)$ are the corresponding single-particle excited
(continuous spectrum) wave-function with the wave-number $n$.
These wave-functions are normalized over the wave-number scale;
see Eq.~(\ref{turbo1}) in Appendix \ref{apa}. This type of
normalization is important for the integration over the
wave-number $K$ in (\ref{gamow}).

The first-order contribution to $\Delta E$ appears to be zero
for $k\to 0^+$ (i.e., for $\mu\to 1^+$). Indeed, we can use (\ref{asa1}) for
\BEA
\label{manuk}
\phi(z_1,z_2)
=\phi(z_1)\,\phi(z_2)=2k\,e^{-k(z_1+z_2)},
\EEA
to conclude
\BEA
&&\langle 0|H|0\rangle=
\int_0^\infty\int_0^\infty\d z_1\,\d z_2
V_{\rm eff}(|z_1-z_2|)\,\phi^2(z_1)\,\phi^2(z_2)\nonumber\\
&&=\int_0^\infty\d v\int_0^v \d u\,
V_{\rm eff}(u)\,
\phi^2\left(\frac{v+u}{2}\right)\,
\phi^2\left(\frac{v-u}{2}\right)=
2k\int_0^\infty\d v e^{-2kv}
\int_0^v \d u\,
V_{\rm eff}(u)={\cal O}(k).
\EEA

Using (\ref{hop1}--\ref{manuk}) we shall calculate various matrix
elements entering into (\ref{gamow}):
\BEA
&&\langle 0|H|n\rangle=\int_0^\infty\int_0^\infty\d z_1\,\d z_2\,
\phi(z_1)\,\phi(z_2)\,V_{\rm eff}(|z_1-z_2|)\,
\phi(z_1)\,\wphi(nz_2, n)\nonumber\\
&&=\sqrt{2k}\,2k\,\int_0^\infty\,\d v\,\int_0^v\d u\,
e^{-k(3v+u)/2}\,V_{\rm eff}(u)\,
\wphi\left(\frac{n(u+v)}{2}, n\right),
\EEA
\BEA
&&\langle 0|H|n_1,\,n_2\rangle
=
\int_0^\infty\int_0^\infty\d z_1\,\d z_2\,
\phi(z_1)\,\phi(z_2)\,V_{\rm eff}(|z_1-z_2|)\,
\wphi(nz_1,n)\,\wphi(nz_2, n)\nonumber\\
&&=2k\,\int_0^\infty\,\d v\,\int_0^v\d u\,
e^{-kv/2}\,V_{\rm eff}(u)\,
\wphi\left(\frac{n_1(u+v)}{2}, n_1\right)
\,\wphi\left(\frac{n_2(v-u)}{2}, n_2\right).
\EEA

This results to the following formula for $\Delta E$,
\BEA
\Delta E=
-2
\int_0^\infty\d n\,\frac{|\langle 0|H|n\rangle|^2}
{\frac{n^2}{2}+\frac{k^2}{2}}
-\int_0^\infty\d n_1\d n_2\,\frac{|\langle 0|H|n_1,\,n_2\rangle|^2}
{\frac{n_1^2}{2}+\frac{n_2^2}{2}+\frac{k^2}{2}}.
\EEA

Working this out and going to the limit $k\to 0^+$ (i.e. $\mu\to 1^+$)
we obtain
\BEA
\label{jk1}
\Delta E&&=-8
\left[
\int_0^\infty\,\d u\,V_{\rm eff}(u)
\right]^2 ~\{~
4\int_0^\infty\frac{\d n}{1+n^2}~\left[
\int_0^\infty\d v\,e^{-3v/2}\wphi\left(\frac{nv}{2},\,0\right)
\right]^2
\\
\label{jk2}
&&+\int_0^\infty\frac{\d n_1\,\d n_2}{2+n_1^2+n_2^2}~\left[
\int_0^\infty\d v\,e^{-v}\wphi\left(\frac{n_1 \,v}{2},\,0\right)\,
\wphi\left(\frac{n_2\, v}{2},\,0\right)
\right]^2
~\}~<~0.
\EEA

This expression for $\Delta E$ is finite in the limit $k\to 0$
(see Appendix \ref{cl} for details), and proportional to the
squared perturbation strength $\left[ \int_0^\infty\,\d u\,V_{\rm
eff}(u) \right]^2$.  In the limit $\mu\to1$ (and for sufficiently
small $\lambda$) we are in the situation where neither
$V(z_1)+V(z_2)$ nor $U$ alone create bound states.  Recalling our
discussion in section \ref{exi} on the existence of bound states
as reflected in the magnitude of variational energy, we conclude
from $\Delta E<0$ that the present approach does predict binding
for $\mu=1$ and for sufficiently small $\lambda$.  Since the
ground state is supposed to be continuous, the very fact of having
a negative energy for $\mu=1$ and not very large $\lambda$ implies
that a bound state will exist for \BEA \label{sumatra}
\lambda_{\rm c}>\lambda>0, \qquad \mu_{\rm c}<\mu<1, \EEA where
neither of the potentials $V$ and $U$ alone allows binding. Here
$\lambda_{\rm c}\geq 1$ is some critical value at which the real
ground-state energy is equal to $E\{ U \}$; recall our discussion
in section \ref{exi}. Note that the precise form of $\xi(\rho)$ is
irrelevant for the argument. This function has to be normalized
and such that the effective potential $V_{\rm eff}$ does not
become large for a sufficiently small $\lambda$ (and, of course,
does not vanish for a finite $\lambda$). For the rest it can be
arbitrary.

Thus in view of (\ref{sumatra}) we have found an example of so called
{\it Borromean binding}, where the involved potentials do not produce bound
states separately, but their cumulative effects lead to such a state.
It is seen from (\ref{ppp}) that this unusual type of binding is
connected with correlations between the $z$-components of each
particles and separately with correlations between their $x$
and $y$ components (which enter via $\rho$).

Note that for three (or more) interacting point-like particles 
(instead of two particles and a
surface) this effect was predicted in nuclear physics; see, e.g.,
Ref.~\cite{Nielsen} for a review.

\section{No-binding conditions}
\label{section_no_binding}

\subsection{First method.}

Here we shall consider certain lower bounds on the sought ground state
energy. Although these bounds are basically algebraic, they are
non-trivial, and they allow to find out under which conditions both the
adsorption and naturation are absent. In this way we complement the
study of the previous section.  We employ |with necessary modifications
and elaborations for our situation| the method suggested in \cite{Fl}.

Note from (\ref{4}) that the presence of the infinite wall can be
modeled via the boundary condition at the plane $z=0$:
\BEA
\label{40}
\Psi(\vec{r}_1,\vec{r}_2)
=0, \quad {\rm if}\quad z_1= 0, \quad
{\rm or}\quad z_2= 0.
\EEA
Though the physical content of the problem demands that
$\Psi(\vec{r}_1,\vec{r}_2)$ is also zero for $z_1<0$ or $z_2<0$, we
can formally require only (\ref{40}) and continue the potential
$V(z)$ to $z<0$ via
\BEA
\label{41}
V(-z)=V (|z|).
\EEA
The ground state energy of the new problem defined with help of
(\ref{40}, \ref{41}) will be obviously equal to the ground state of the original
problem.

Let us now introduce a fictive particle with the mass $M$ and the
radius vector
\BEA
\vec{r}_3=(x_3,y_3,z_3).
\EEA
Now Eq.~(\ref{41}) is generalized
to the corresponding translation-invariant interaction with the
fictive particle:
\BEA
\label{42}
V(|z_k-z_3|), \qquad k=1,2.
\EEA
It is again obvious that upon taking the limit $M\to \infty$, the
motion of the fictive particle will completely freeze, $\vec{r}_3$ will
reduce to a constant which can be taken equal to zero.

Thus the three-particle (two real particles plus the fictive one)
Schr\"odinger equation reads analogously to (\ref{1a}, \ref{2a})
\BEA
\label{100}
\left \{
-\frac{1}{2M}\,\frac{\partial^2}{\partial \vec{r_3}^2}
-\frac{1}{2}\,\frac{\partial^2}{\partial \vec{r_1}^2}-\frac{1}{2}
\,\frac{\partial^2}{\partial \vec{r_2}^2}+
V(|z_1-z_3|)+V(|z_2-z_3|)+U(|\vec{r}_1-\vec{r}_2|)
-E(M)\right\}\Psi
=0,
\EEA
the correct two-particle energy being recovered in the limit $M\to\infty$.

Note that the boundary conditions (\ref{40}) are modified as well
\BEA
\label{400}
\Psi=0, \quad {\rm if}\quad z_1= z_3, \quad
{\rm or}\quad z_2= z_3.
\EEA

It is seen that the Hamiltonian in (\ref{100}) is invariant with respect to
simultaneous shift of all three radius vectors $\vec{r}_k$ ($k=1,2,3$) by
some vector. Since we consider a finite-particle quantum system,
symmetry of the Hamiltonian implies the symmetry of the corresponding
ground-state wave-function. Thus we deduce for this function
\BEA
\Psi=\Psi(\vec{r_1}-\vec{r_2},\,\vec{r_1}-\vec{r_3},\,\vec{r_2}-\vec{r_3}),
\EEA
which implies
\BEA
\label{1000}
\left \{
\frac{\partial}{\partial \vec{r_3}}
+\frac{\partial}{\partial \vec{r_1}}+
\,\frac{\partial}{\partial \vec{r_2}}
\right\}\Psi
=0.
\EEA

We shall now decompose the Hamiltonian in
(\ref{100}) such that (\ref{1000}) is employed and that the separate
sectors of the problem |i.e., surface-particle and inter-particle
interaction| are made transparent:
\BEA
\label{t0}
&&H\equiv
-\frac{1}{2M}\,\frac{\partial^2}{\partial \vec{r_3}^2}
-\frac{1}{2}\,\frac{\partial^2}{\partial \vec{r_1}^2}
-\frac{1}{2}\,\frac{\partial^2}{\partial \vec{r_2}^2}
+V(|z_1-z_3|)+V(|z_2-z_3|)+U(|\vec{r}_1-\vec{r}_2|),\\
&&=H_{0}+H_{12}+H_{13}+H_{23},\\
\label{t1}
&&H_{0}\equiv
-\frac{1}{2}\left(
\frac{\partial}{\partial \vec{r_3}}
+\frac{\partial}{\partial \vec{r_1}}+
\frac{\partial}{\partial \vec{r_2}}
\right)
\left(
a\,\frac{\partial}{\partial \vec{r_3}}
+b\,\frac{\partial}{\partial \vec{r_1}}+
b\,\frac{\partial}{\partial \vec{r_2}}
\right)
\\
\label{t2}
&&H_{13}\equiv
-\frac{c}{2}\left(
\frac{1}{1+x}\frac{\partial}{\partial \vec{r_1}}-
\frac{x}{1+x}\frac{\partial}{\partial \vec{r_3}}
\right)^2
+V(|z_1-z_3|)
\\
\label{t3}
&&H_{23}\equiv
-\frac{c}{2}\left(
\frac{1}{1+x}\frac{\partial}{\partial \vec{r_2}}-
\frac{x}{1+x}\frac{\partial}{\partial \vec{r_3}}
\right)^2
+V(|z_2-z_3|)
\\
\label{t4}
&&H_{12}\equiv
-2d\left(\frac{1}{2}\,
\frac{\partial}{\partial \vec{r_1}}-
\frac{1}{2}\,
\frac{\partial}{\partial \vec{r_2}}
\right)^2+U(|\vec{r}_1-\vec{r}_2|).
\EEA

The coefficients $a,b,c$ and $d$ are read off directly from (\ref{t0}--\ref{t4}):
\BEA
&&a=-\frac{2x^2}{(1+2x)^2},\\
\label{kabo1}
&&c=\frac{(1+x)^2}{(1+2x)^2},\\
&&b=d=\frac{2x(1+x)}{(1+2x)^2},
\label{kabo2}
\EEA
where the limit $M\to\infty$ has already been taken.
Here $x$ is a free parameter; the boundaries of its change
are to be determined below.

Let us now take average of the Hamiltonian $H$ with the ground-state
wave function $\Psi$. The term $\langle\Psi|H_0|\Psi\rangle$ is zero
due to (\ref{1000}). We shall now establish when the remaining terms
in $\langle\Psi|H|\Psi\rangle$ are certainly positive, that is when
bound|i.e., naturated {\it or} adsorbed| states are certainly absent.

Changing the variables as
\BEA
\vec{\xi}_{1}
=(1+2x)\vec{r}_{1}+\vec{r}_{3},
\qquad
\vec{\xi}_{3}
=x\vec{r}_{1}+\vec{r}_{3},
\EEA
one reduces $H_{13}$ to a form
\BEA
H_{13}=
-\frac{c}{2}\frac{\partial^2}{\partial \vec{\xi}_{1}^{\,2}}
+V\left(\,\left|{\xi}_{1z}-2 {\xi}_{3z}\right|\,\right),
\EEA
where $\xi_{1z}$ and $\xi_{3z}$ are the third components of the vectors
$\vec{\xi}_{1}$ and $\vec{\xi}_{3}$, respectively.
The constant factor $2\xi_{3z}$ will obviously not
change the binding conditions. Recalling boundary conditions (\ref{400})
we see that $\langle\Psi|H_{13}|\Psi\rangle$ is certainly
positive for
\BEA
\label{krim}
\mu\leq c,
\EEA
where $\mu$ is the coupling constant of $V$, such that $H_{13}$ with $c=1$
has the binding threshold $\mu=1$ [compare with (\ref{petan}, \ref{dur})]. Obviously,
$\langle\Psi|H_{23}|\Psi\rangle$ is positive under the same condition
(\ref{krim}).

As for $\langle\Psi|H_{12}|\Psi\rangle$ we change the variables as
\BEA
\label{lisuk}
\vec{r}_{12}
={\vec{r}_{1}-\vec{r}_{2}},
\qquad
\vec{R}_{12}
=\frac{\vec{r}_{1}+\vec{r}_{2}}{2},
\EEA
to see that $H_{12}$ takes the form
\BEA
H_{13}=
-\frac{2d}{m}\frac{\partial^2}{\partial \vec{r}_{12}^2}
+U(r_{12}).
\EEA
Thus, $\langle\Psi|H_{13}|\Psi\rangle$ is certainly positive for
\BEA
\label{kriminal}
\lambda\leq 2d,
\EEA
where $\lambda$ is the coupling constant of $U$, such that $H_{12}$ with $2d=1$
has the critical binding threshold $\lambda=1$ [compare with (\ref{mukhan},\ref{baran})].

Let us now recall that we employed $c$ and $2d$ as inverse effective
masses which should be positive; thus we should restrict
ourselves to the situations
\BEA
\label{vesna}
x\geq 0
\EEA
and $x<-1$, as seen from (\ref{kabo1}, \ref{kabo2}). As inspection shows, the relevant no-binding
condition is produced for $x$ changing from zero to plus infinity, i.e., for
the branch (\ref{vesna}).

Thus, under conditions (\ref{krim}, \ref{kriminal}),
where the limit $M\to\infty$ is being taken, the overall bound states
are certainly absent.

\subsection{Second method.}

Let us now turn to another, simpler way of deriving
no-binding regions. For some range of parameters the
present method will have a priority over the considered one, and
then by combining the two methods we shall get an extended
no-binding region. We return to the very original quantum
Hamiltonian in (\ref{2a}) and write it as \BEA \label{karusel1}
&&-\frac{\alpha}{4}\,\left( \frac{\partial}{\partial \vec{r_1}}
+\frac{\partial}{\partial \vec{r_2}}
\right)^2\\
\label{karusel2}
&&-\frac{\alpha}{4}\,\left(
\frac{\partial}{\partial \vec{r_1}}
-\frac{\partial}{\partial \vec{r_2}}
\right)^2+U(|\vec{r}_1-\vec{r}_2|)\\
\label{karusel3}
&&
-\frac{1-\alpha}{2}\,\frac{\partial^2}{\partial \vec{r_1}^2}+V(\vec{r}_1)\\
&&-\frac{1-\alpha}{2}\,\frac{\partial^2}{\partial \vec{r_2}^2}+
V(\vec{r}_2),
\label{karusel4}
\EEA
where
\BEA
0\leq \alpha\leq 1.
\label{leto}
\EEA

The term in (\ref{karusel1}) is seen to be always positive; for the
term (\ref{karusel2}) we change the variables as in (\ref{lisuk}), to
get that it is always positive for
\BEA
\label{111}
\lambda\leq\alpha,
\EEA
while the terms in (\ref{karusel3}, \ref{karusel4}) are both positive
under
\BEA
\label{222}
\mu \leq 1-\alpha.
\EEA
Here $\lambda$ and $\mu$ have the same meaning as in
(\ref{krim}, \ref{kriminal}). Thus no binding is possible
if (\ref{111}) and (\ref{222}) are satisfied simultaneously. 

\subsection{Convexity argument.}

As for the last ingredient of our construction, we note that the
coupling constants $\mu$ and $\lambda$ enter into Hamiltonian
$H(\mu,\lambda)$ in the linear way, and that the following
convexity feature is valid for the ground state as a function of
$\mu$ and $\lambda$: \BEA \label{co1} &&{\rm min}\,\left[\, H(\nu
\mu_1+(1-\nu) \mu_2,\, \nu \lambda_1+(1-\nu) \lambda_2 )\,\right]
\\
&&=
{\rm min}\,\left[\nu H( \mu_1,\lambda_1)
+(1-\nu) H( \mu_2,\lambda_2)\right]
\geq
\nu\,{\rm min}\,\left[H( \mu_1,\lambda_1)\right]+
(1-\nu)\,{\rm min}\,\left[H( \mu_2,\lambda_2)\right].
\label{co2}
\EEA

In other words, if in the phase diagram the binding|i.e., naturation {\it or} adsorption|
is prohibited at
points $( \mu_1,\lambda_1)$ and $( \mu_2,\lambda_2)$ |that is ${\rm
  min}\,\left[H( \mu_1,\lambda_1)\right]\geq 0$ and ${\rm
  min}\,\left[H( \mu_2,\lambda_2)\right]\geq 0$| then there is no
binding on the whole line connecting those two points, because from
(\ref{co1}, \ref{co2}) one has ${\rm min}\,\left[\, H(\nu
  \mu_1+(1-\nu) \mu_2,\, \nu \lambda_1+(1-\nu) \lambda_2 )\,\right]\geq
0$.

Thus we draw together the bounds (\ref{krim}, \ref{kriminal},
\ref{111}, \ref{222}) |under conditions (\ref{vesna}, \ref{leto})
determining the ranges of the parameters $x$ and $\alpha$,
respectively| and complete it to a convex figure ensuring
that for every two points belonging to (\ref{krim}, \ref{kriminal},
\ref{111}, \ref{222}) the line joining them is also considered as
binding-prohibited. The result is presented in Fig. 1. It is seen that
there is the critical strength $\mu_c=0.25$| which is necessary for
binding.

The latter value of $\mu$ is special for the following reason: for
$\lambda\to\infty$, i.e., when the inter-particle attraction is too
strong, the two particles are tightly connected to each other.  The mass
of the composite particle is two times larger, and (at the same time)
the potential acting on it is two times larger. This leads to the
adsorption threshold $\mu=0.25$, which is independently obtained via the
above no-binding conditions.

\section{Phase diagram.}

We are now prepared to present in Fig. 1 the qualitative phase diagram
of the model. The axes of the phase diagram are $\lambda$ and $\mu$. The
dimensionless parameter $\lambda$ enters into the inter-strand
interaction energy $\lambda\widetilde{U}$, such that without the
adsorbing surface the naturated phase of the two strands exists only for
$\lambda\geq 1$; see sections \ref{urchin2} and \ref{urchin3} for
details.  In this phase the two strands are localized next to each other
and their fluctuations are correlated. The typical form of $\lambda$ for
the considered short-range potentials is
\BEA
\label{bumerang}
\lambda=\frac{c\nu}{a^2l_p^2T},
\EEA
where $c$ is a numerical prefactor, $\nu$ is the strength of the
inter-strand potential (i.e., the modulus of its minimal value), $l_p$
is the persistence length, and $T$ is temperature (recall that
Boltzmann's constant is unity, $k_B=1$); see sections \ref{urchin2} for
mode details. In particular, recall that for the Morse potential
discussed around Eq.~(\ref{eth2}) the concrete formula for $\lambda$
reads $\lambda=\frac{16\nu}{a^2l_p^2T}$, where $\nu\simeq 0.01$eV,
$l_p\simeq 1$nm and $a\,l_p\simeq$2 \cite{PBreview}.  Taking room
temperatures for $T$ we get that $\lambda\sim 1$. 

Analogously, the dimensionless parameter $\mu$ enters the strand-surface
attractive potential as $\mu\widetilde{V}$, such that the adsorbed phase
of one single strand (that is without inter-strand interaction) exists
for $\mu\geq 1$; see sections \ref{urchin1} and \ref{urchin3} for
details. Note that $\mu$ has the same qualitative form (\ref{bumerang}),
where $\nu$ is the strength of the surface-strand potential. Recall that
for the electrostatic surface-monomer attraction the concrete expression
for $\mu$ is discussed in (\ref{eth1}): $\mu=\frac{8.33\, \pi |\sigma
q|}{T\,k_{\rm D}^3 l_p^2 \epsilon}$,.  where $k_{\rm D}^{-1}$ is the
Debye screening length ($k_{\rm D}^{-1}=0.1$nm at normal conditions),
$q$ is the monomer charge (around one electron charge for a
single-strand DNA), $l_p$ is the persistence legth (around $1$nm for a
single-strand DNA), and finally $\sigma$ is the charge density of the
surface. Strongly charged surfaces have typically $1-10$ electron
charges per $1\,{\rm nm}^2$. At room temperatures $\mu\sim 1$. 

In Fig. 1 the thermodynamical phases are confined by
thick lines.  ${\bf ND}$, ${\bf NA}$ and ${\bf DD}$ refer, respectively,
to the naturated-desorbed, naturated-adsorbed, and denaturated-desorbed
phases. The meaning of these term should be self-explanatory, e.g., in
the ${\bf ND}$ phase the two strands are localized next to each other,
but they are far from the surface.

First of all we see that there is no adsorbed and denaturated phase: as
we have shown already in section \ref{section_no_melting}, even small
(but generic) inter-strand (inter-particle) attraction suffices to
create a naturated state, provided that the two strands (particles) are
adsorbed. Thus the rectangular region {\bf c} in Fig. 1, which belongs
to the naturated and adsorbed phase {\bf NA}, refers to conditions where
the overall binding is due to sufficiently strong attraction to the
surface. 

The curved line going from $(\mu=1,\,\lambda=0)$ to
$(\mu=0.25,\,\lambda=1)$ in Fig. 1 confines region {\bf a}, where no
overall binding (i.e., no-denaturation and no-adsorption) is possible
according to the lower bounds obtained in the previous section. 

The region {\bf b}, confined by two straight normal lines and the thick
curve, refers to the the collective binding situation. It is seen that
this region lies below both adsorption and denaturation thresholds.
While we do not know the precise position of the thick curve confining
the region {\bf b}, we proved its existence in section
\ref{section_borromean_binding}.

Finally, the line separating {\bf NA} (naturated-adsorbed) phase from
{\bf ND} (naturated-desorbed) phase extends monotonically to $\mu=0.25$
for $\lambda\to \infty$. Please note that the monotonicity of this line
is conjectured.  Still this conjecture is, to our opinion, quite likely
to be correct.

\begin{figure}
\includegraphics[width=0.35\linewidth]{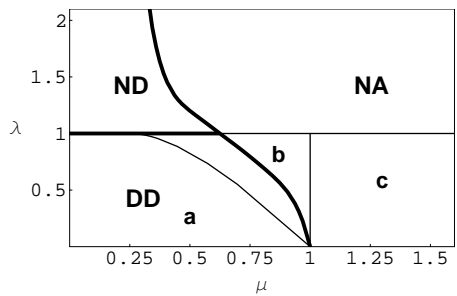}
\hfill
\caption{
Schematic phase diagram for
the inter-strand coupling  $\lambda$ versus
 the strand-surface coupling $\mu$. The bold lines confine
three thermodynamical phases.
{\bf ND}: Naturation and desorption.
{\bf NA}: Naturation and adsorption.
{\bf DD}: Desorption and denaturation.
The critical naturation strength in the
bulk is $\lambda_c=1$, for single strand adsorption it is $\mu_c=1$.
The following subregions are confined by normal lines.
{\bf a}: Domain described by the no-binding condition  of section \ref{section_no_binding}.
{\bf b} (bounded  by the bold {\bf DD}-{\bf NA} line and two straight segments):
Borromean naturation and adsorption.
{\bf c}: Adsorption and naturation due to overcritical
coupling to the  surface.
}
\label{1f}
\end{figure}

\section{Summary.}

The main purpose of this paper was in studying DNA denaturation in
the presence of an adsorbing plane surface.  As we argued in the
introduction, there are several relevant situations when the two
processes, adsorption and denaturation, are encountered together.
Taking into account the importance of these processes
in the physics of DNA, as well as for DNA-based technologies, it is
important to understand how specifically adsorption and
denaturation interact with each other.

Our two basic findings can be summarized as follows. First we saw that
provided the two strands of DNA are (even weakly) adsorbed on the surface, there is no
denaturation phase transition. There is only a smooth crossover from the
naturated state to a (very) weakly bound state.  Second we have shown
that when the inter-strand attraction alone and the surface-strand
attraction alone are too weak to create naturated and adsorbed state,
respectively, their combined effect (``Borromean binding'') can create
such a naturated and adsorbed state. 

The results were displayed on a simple model of two coupled homopolymers
(strands) interacting with the plane surface. The volume interaction within each homopolymer
can be accounted for via renormalizing the persistence length; see section \ref{tartar}.
Many realistic features of DNA are thereby put aside; see the beginning of section
\ref{section_model}. We plan to investigate some of them elsewhere. Another interesting subject
is to study the DNA adsorption on a curved surface \cite{winkler,tsai}. 

We, nevertheless, hope that the basic qualitative aspects of the presented problem
are caught adequately, and that the presented results increase our
understanding of DNA physics. 

\acknowledgments

The authors thank the unknown referees for their constructive remarks.

This work was supported by the National Science
Council of the Republic of China (Taiwan) under Grant No. NSC
95-2112-M 001-008, National Center of Theoretical Sciences in
Taiwan, and Academia Sinica (Taiwan) under Grant No. AS-95-TP-A07.

A.E. A. was supported by Volkswagenstiftung and by CRDF Grant No. ARP2-2647-YE-05.

\appendix

\newpage

\section{Derivation of the Schr\"odinger equation from transfer matrix.}
\label{effective_wave_equation}

Imposing the periodic boundary conditions, the partition function
(\ref{glum2}) can be written as \BEA {\cal Z}={\rm Tr}\, {\cal
T}^N\equiv \int\d\vec{r}_{1}\, \d\vec{r}_{2} \, {\cal
T}^N(\vec{r}_{1}, \vec{r}_{2}; \vec{r}_{1}, \vec{r}_{2}), \EEA
where ${\cal T}$ is the transfer operator parametrized with two
continuous indices: 
\BEA \label{cobalt_1} {\cal T}(\vec{r}_{1},
\vec{r}_{2}; \vec{r}^{\,'}_{1}, \vec{r}^{\,'}_{2})= \exp\left[
-\beta({\cal U}(\vec{r}_{1}- \vec{r}_{2})+{\cal V}(\vec{r}_{1})+
{\cal V}(\vec{r}_{2}) )-\frac{\beta \l}{2}(\vec{r}_{1}- \vec{r}^{\,
'}_{1})^2- \frac{\beta \l}{2}(\vec{r}_{2}- \vec{r}^{\, '}_{2})^2
\right]. 
\EEA

Thus in the thermodynamic limit $N\to \infty$:
\BEA
\label{boreni}
{\cal Z}=\Lambda^N,
\EEA
where $\Lambda$ is the largest eigenvalue of ${\cal T}$.

For simplicity reasons, the subsequent discussion will be done in terms
of a transfer matrix, which depends on a two scalar variables $z'$ and $z$. The extension
to the more general case (\ref{cobalt_1}) is straightforward.

Write the eigenvalue equation for the right eigenvector as
\BEA
\label{lala}
\int \d z'\, e^{-\beta \CV(z) -\frac{\beta \l}{2}(z- z')^2 }\psi (z')=e^{-\beta f }\psi(z),
\EEA
where $e^{-\beta f }$ and $\psi$ are, respectively,
eigenvalue and eigenvector. It is seen from (\ref{boreni}) that $Nf$ is the free energy of the 
model in the thermodynamic limit $N\gg 1$ provided that there is a gap between the largest eigenvalue
$\Lambda$ and the one but largest eigenvalue.

One now assumes that
\BEA
\label{kaban}
\beta \l D^2
\gg 1,
\EEA
where $D$ is the characteristic length of $\psi(z)$. 
Since $\psi^2(z)$ is the density of monomers,
we see that $D$ quantifies the thickness of the adsorbed layer. 
Recalling (\ref{LK}) we can write condition (\ref{kaban}) as
\BEA
\label{varaz}
D\gg l_p,
\EEA
i.e., the thickness is much larger than the persistence length.

Under condition (\ref{kaban}) the dominant part of the integration in (\ref{lala}) is $z\approx z'$.
With this in mind we expand $\psi(z')$ in (\ref{lala}) as
\BEA
\psi(z')=\psi(z)+(z-z')\psi'(z)+\frac{(z-z')^2}{2}\psi''(z)+...,
\EEA
and substitute this expansion into (\ref{lala}). The outcome is
\BEA
\label{lala1}
\frac{\sqrt{2\pi}}{\sqrt{\l\beta}}
e^{-\beta \CV(z)}\left(
1+\frac{1}{2\beta \l}\frac{\d^2}{\d z^2}
\right)
\psi (z)=e^{-\beta f }\psi(z).
\label{peto}
\EEA
The corrections to this equation are of order $O(\frac{1}{\l^2\beta^2 D^4})=
O(\frac{l_p^4}{D^4})$.

Eq.~(\ref{lala1}) can be re-written as
\BEA
\label{ssq}
\frac{1}{2\beta \l}\frac{\d^2}{\d z^2}\psi(z)
=\left[ e^{\beta (\, \CV(z) - \widetilde{f} \,) } - 1 \right]\psi(z), \qquad
\widetilde{f}\equiv f+\frac{T}{2}\ln \frac{2\pi}{K\beta}.
\EEA

For weakly-bound states
\BEA
|\CV(z) - \widetilde{f}|\ll 1,
\EEA
for those $z$, where $|\psi(z)|$ is sufficiently far from zero. Thus in (\ref{ssq}) we can expand 
\BEA
\label{tan}
e^{\beta (\, \CV(z) - \widetilde{f} \,) } - 1 \simeq \beta (\, \CV(z) - \widetilde{f} \,)
\EEA
and get the Schr\"odinger equation:
\BEA
\label{erwin_a}
\left(-\frac{1}{2}\frac{\d^2}{\d z^2}+
\beta^2 \l \CV(z)
\right)\psi(z)
=\left( \beta^2 \l f + \frac{\beta \l}{2} \ln  \frac{2\pi}{\beta \l}            \right)\psi(z)
\equiv E\psi(z).
\EEA

The ground-state energy $E$ of this Schr\"odinger equation relates to the free energy $f$ of the
original polymer problem. For weakly-bound states $E$ is negative and close to zero. 

\section{Solution of Schr\"odinger equation with the delta-shell potential.}
\label{apa}
\subsection{Discrete spectrum.}

Here we outline bound-state solutions of
a one-dimensional Schr\"odinger equation
\BEA
-\frac{1}{2m}\psi ''(x)+(V(x)-E)\psi(x)=0
\label{bud}
\EEA
with the attractive delta-shell potential
\footnote{
We should like to clarify the physical meaning of studying the
delta-shell potential (\ref{kuma}). First of all it should be clear that
the weak-potential condition (\ref{tan}) does not (formally) hold for
the strongly singular potential (\ref{kuma}). Thus the transition from
the transfer-matrix equation to the Schr\"odinger equation is {\it
formally} not legitimate.  Nevertheless, there is a clear reason for
studying the potential (\ref{kuma}) in the context of polymer physics,
since it is known that the physics of weakly-bound quantum particles in
a short-range binding potential does not depend on details of this
potential \cite{LL}.  So once the conditions for going from the
transfer-matrix equation to the Schr\"odinger equation are satisfied for
some short-range potential, one can employ the singular potential
(\ref{kuma}) for modeling features of weakly-bound particles in that
potential. This is in fact the standard idea of using singular
potentials. }
\BEA
\label{kuma}
V(x)=-\frac{\mu}{2mx_0}\,\delta(x-x_0),
\EEA
and with the infinite wall at $x=0$:
\BEA
\label{korund}
\psi(0)=0.
\EEA
Here $\lambda>0$ is the dimensionless
strength of the potential, while $x_0$ is the radius of attraction.
$m$ is the particle mass. In contrast to the main text, here we do not
put $m=1$.

Due to boundary condition (\ref{korund}) the considered problem is equivalent to
the corresponding 
three-dimensional Schr\"odinger problem with centrally-symmetric potential. 

Let us rewrite (\ref{bud}) as
\BEA
\psi''(x)-k^2\psi(x)=-\frac{\mu}{x_0}\,\delta(x-x_0)\,\psi,
\qquad k\equiv \sqrt{2m|E|}\geq 0.
\label{budo}
\EEA
For $x\not =x_0$ (\ref{budo}) is a free wave-equation. Its solution
for $x < x_0$ and $x > x_0$ are found from the boundary conditions
$\psi(x=0)=0$ and $\psi(x\to \infty)=0$, respectively.
Thus the overall solution is obtained as
\BEA
\label{turbo}
&&\psi(x)={\cal N}^{-1/2}\, \sinh (kx_<)\,e^{-kx_>},\\
&&x_<\equiv {\rm min}\,(x,x_0),\qquad
x_>\equiv {\rm max}\,(x,x_0),
\EEA
where ${\cal N}$ is the normalization constant
determined via $\int_0^\infty\d x\, \psi^2(x)=1$,
\BEA
\label{norma}
{\cal N}=\frac{e^{-2kx_0}}{4k}\,\left[
\sinh (2kx_0)-2kx_0
\right]+\frac{\sinh^2(kx_0)}{2k}\,e^{-2kx_0}.
\EEA

Substituting (\ref{turbo}) into (\ref{budo}) we get an equation
for the energy of the single bound-state: \BEA \label{koza}
kx_0=\mu\,\sinh (kx_0)\,e^{-kx_0}. \EEA The critical strength of
the potential is seen to be \BEA \label{crit} \mu=1, \EEA because
for small $kx_0$ Eq.~(\ref{koza}) gives \BEA \label{gamala}
kx_0=\frac{\mu-1}{\mu}. \EEA

Note the following form of $\psi(x)$ for small values of $kx_0$:
\BEA
\label{asa}
\psi(x)= \frac{\sqrt{2k}\,x_<}{x_0}\,e^{-k(x_>-x_0)}+{\cal O}(k).
\EEA

For large values of $\lambda$ the bound state energy increases as
\BEA
2kx_0=\mu.
\EEA

\subsection{Continuous spectrum.}

For studying the continuous (positive-energy)
spectrum of Eq.~(\ref{bud}), we re-write it as
\BEA
\wpsi''(x)+\tk^2\wpsi(x)=-\frac{\mu}{x_0}\,\delta(x-x_0)\,\wpsi,
\qquad \tk\equiv \sqrt{2mE}\geq 0.
\label{budo1}
\EEA

The solution is found as
\BEA
\label{turbo1}
\wpsi(\tk x,\tk)
=\frac{\sqrt{2}}{\sqrt{\pi}\,\sin(\tk\,x_0)}
\, \sin(\tk x_<) \sin( \tk x_> +\delta(\tk)\,)
\EEA
where $\delta(\tk)$ is the phase-shift to be determined below, and
where $\widetilde{{\cal N}}$ is the normalization constant
determined via orthogonalization on the $\tk$-scale:
\BEA
\int_0^\infty\d x\, \wpsi(\tk\,x,\tk)~\wpsi(\tk\,x,\tk ')
=\delta(\tk-\tk ').
\EEA

This normalization can be checked via the large-$x$ behavior of
$\wpsi(x,\tk)$ \cite{LL}. Note that for $\wpsi(\tk x,\tk) $ there
are two types of dependence on the wave-number $n$: as a
prefactor for the argument and as a parameter entering into the
normalization and the phase-shift.

For the phase-shift $\delta(\tk)$ we get:
\BEA
\tk \,x_0\sin\delta(\tk)=\mu \sin(\tk\,x_0)\sin(\tk\,x_0+\delta(\tk)\,),
\EEA
that for $\tk\to 0$ reduces to
\BEA
\frac{1}{\mu}-1=\frac{\tk\,x_0}{2}\cot\left(\delta(\tk)\,\right).
\EEA
Thus
\BEA
\delta(0)=0,
\EEA
and for $\wpsi(\tk x,0)$ we have
\BEA
\label{turbo11}
\wpsi(\tk x,0)
=\frac{\sqrt{2}\,x_<}{\sqrt{\pi}\,x_0}
\,\sin( \tk x_>).
\EEA

\section{}
\label{cl}

While the finiteness of the integral (\ref{jk1}) is obvious
(because the integral $\int_0^\infty\d n/(1+n^2)$ is already convergent),
the convergence of the integral in (\ref{jk2}) is less
trivial. Estimating from (\ref{turbo1}, \ref{turbo11})
\BEA
\wphi(nx,0)=\sqrt{\frac{2}{\pi}} \sin (nx),
\EEA
we get for the integral in (\ref{jk2}):
\BEA \frac{4}{\pi^2}
\int_0^\infty\frac{\d n_1\,\d n_2}{\frac{1}{2}+n_1^2+n_2^2}~
\frac{n_1^2n_2^2}{\left[
(n_1^2-n_2^2)^2
+2(n_1^2+n_2^2)+1
\right]^2}
=\frac{1}{4\pi^2}\,
\int_0^{2\pi}\int_0^\infty
\frac{\d \alpha\, n\,\d n}{\half+n^2}~\frac{n^4\sin^22\alpha}{\left[
n^4\cos^22\alpha+2n^2+1
\right]^2}\\
=\frac{1}{4\pi^2}\,
\int_0^{\pi}\int_0^\infty
\frac{\d \alpha\,\d n}{\half+n}~\frac{n^2\sin^2\alpha}{\left[
n^2\cos^2\alpha+2n+1
\right]^2}.
\label{koop1}
\EEA

The integral over $n$ in (\ref{koop1}) is convergent and produces an
integrable logarithmic singularity $\sim\ln \cos^2\alpha$.  Thus the
double integral in (\ref{koop1}) is finite.

\end{document}